\documentclass[modern]{aastex631}

\newcommand{\Htwo}{\(\textrm{H}_{2}\)}

\newcommand{\CHthCN}{\(\textrm{CH}_{3}\textrm{CN}\)}
\newcommand{\HCthN}{\(\textrm{HC}_{3}\textrm{N}\)}

\begin{document}

\title{UV-driven Chemistry as a Signpost for Late-stage Planet Formation}

\author{Jenny K. Calahan}
\affiliation{University of Michigan, 323 West Hall, 1085 South University Avenue, Ann Arbor, MI 48109, USA}

\author{Edwin A. Bergin}
\affiliation{University of Michigan, 323 West Hall, 1085 South University Avenue, Ann Arbor, MI 48109, USA}

\author{Arthur D. Bosman}
\affiliation{University of Michigan, 323 West Hall, 1085 South University Avenue, Ann Arbor, MI 48109, USA}

\author{Evan Rich}
\affiliation{University of Michigan, 323 West Hall, 1085 South University Avenue, Ann Arbor, MI 48109, USA}

\author{Sean M. Andrews}
\affiliation{Center for Astrophysics \textbar Harvard \& Smithsonian, 60 Garden St., Cambridge, MA 02138, USA}

\author{Jennifer B. Bergner}
\affiliation{University of Chicago Department of the Geophysical Sciences, Chicago, IL 60637, USA}
\altaffiliation{NASA Hubble Fellowship Program Sagan Fellow}

\author{L. Ilsedore Cleeves}
\affiliation{Department of Astronomy, University of Virginia, Charlottesville, VA 22904, USA}

\author{Viviana V. Guzm\'an}
\affiliation{Instituto de Astrofísica, Ponticia Universidad Católica de Chile, Av. Vicuña Mackenna 4860, 7820436 Macul, Santiago, Chile}
\affiliation{N\'ucleo Milenio de Formaci\'on Planetaria (NPF), Chile}

\author{Jane Huang}
\affiliation{University of Michigan, 323 West Hall, 1085 South University Avenue, Ann Arbor, MI 48109, USA}
\altaffiliation{NASA Hubble Fellowship Program Sagan Fellow}

\author{{John~D.~Ilee}}
\affiliation{School of Physics \& Astronomy, University of Leeds, Leeds LS2 9JT, UK}

\author{Charles J. Law}
\affiliation{Center for Astrophysics \textbar Harvard \& Smithsonian, 60 Garden St., Cambridge, MA 02138, USA}

\author{Romane Le Gal}
\affiliation{Center for Astrophysics \textbar Harvard \& Smithsonian, 60 Garden St., Cambridge, MA 02138, USA}
\affiliation{IRAP, Universit\'{e} de Toulouse, CNRS, CNES, UT3, 31400 Toulouse, France}
\affiliation{IPAG, Universit\'{e} Grenoble Alpes, CNRS, IPAG, F-38000 Grenoble, France}
\affiliation{IRAM, 300 rue de la piscine, F-38406 Saint-Martin d'H\`{e}res, France}

\author{Karin I. \"Oberg}
\affiliation{Center for Astrophysics \textbar Harvard \& Smithsonian, 60 Garden St., Cambridge, MA 02138, USA}

\author{Richard Teague}
\affiliation{Center for Astrophysics \textbar Harvard \& Smithsonian, 60 Garden St., Cambridge, MA 02138, USA}
\affiliation{Department of Earth, Atmospheric, and Planetary Sciences, Massachusetts Institute of Technology, Cambridge, MA 02139, USA}

\author{Catherine Walsh}
\affiliation{School of Physics \& Astronomy, University of Leeds, Leeds LS2 9JT, UK}

\author{David J. Wilner}
\affiliation{Center for Astrophysics \textbar Harvard \& Smithsonian, 60 Garden St., Cambridge, MA 02138, USA}

\author{Ke Zhang}
\affiliation{Department of Astronomy, University of Wisconsin-Madison, 475 N Charter St, Madison, WI 53706}

\begin{abstract}

The chemical reservoir within protoplanetary disks has a direct impact on planetary compositions and the potential for life. A long-lived carbon-and nitrogen-rich chemistry at cold temperatures ($\leq$
50 K) is observed within cold and evolved planet-forming disks. This is evidenced by bright emission from small organic radicals in 1–10 Myr aged systems that would otherwise have frozen out onto grains within 1 Myr. We explain how the chemistry of a planet-forming disk evolves from a cosmic-ray/X-ray-dominated regime to a ultraviolet-dominated chemical equilibrium. This,in turn, will bring about a temporal transition in the chemical reservoir from which planets will accrete. This photochemical dominated gas phase chemistry develops as dust evolves via growth, settling and drift, and the small grain population is depleted from the disk atmosphere. A higher gas-to-dust mass ratio allows for deeper penetration of ultraviolet photons is coupled with a carbon-rich gas (C/O $>$ 1) to form carbon-bearing radicals and ions. This further results in gas phase formation of organic molecules, which then would be accreted by any actively forming planets present in the evolved disk.

\end{abstract}

\section*{\large Main} \label{sec:intro}

Protoplanetary disks are the natal environments for planets. Disks have three main components: a pebble-rich dusty midplane (dust grain radius $>\sim$ 1mm), a gaseous atmosphere extending well above\citep{Miotello22} and radially beyond (by a factor of $\sim$2\citep{Ansdell18}) the pebble-rich midplane, and a small dust population (radius $<$ 10 $\mu$m) that is coupled to the gas. Each component of the protoplanetary disk has an impact on shaping the chemistry of actively forming planets. The solid cores of giant planets must form over a short timescale ($\sim$1~Myr) for the eventual planet to obtain its full mass over the course of a typical lifetime of gas in a disk (3-10~Myr\citep{Haisch01}) and to explain the widely observed gap and ring structures that are thought to be indicative of planet formation\citep{Andrews18}. The compositions of pebbles and their icy mantles directly influence the final composition of a solid planetary core \citep{Oberg11ice,Johansen17}. After a core becomes sufficiently massive, planets start to accrete material from the gaseous reservoir surrounding the pebble-rich midplane of a protoplanetary disk to form their atmospheres \citep{Lambrechts14}. It remains difficult to directly probe the gas within the planet-forming midplane due to the high dust densities leading to elevated dust optical depths that mask line emission, as well as cold temperatures which leads to the freezing out common gas tracers such as CO onto dust grains. However, constraining the chemical environment of the planet-forming midplane is essential to connect sub-mm observations probing the warm intermediate regions above the disk midplane to the composition of actively forming planets. 

The molecules \CHthCN{} and \HCthN{} are two of many complex organic molecules (loosely defined as a molecules with at least four atoms, including multiple carbon atoms) that could act as basic precursors to prebiotic molecules\citep{Powner09, Ritson12, Sutherland15}. \CHthCN{} and \HCthN{} have been observed and spatially resolved towards the protoplanetary disks around six young stars: GM Aur, AS 209, HD 163296, MWC 480, LkCa 15, and V4046 Sgr \citep{Bergner18, Ilee21, Kastner18,Oberg15}. \CHthCN{} has also been spatially resolved observed toward the TW Hya disk\citep{Loomis18}, and a couple of small ($\approx$4 source) surveys have detected unresolved \CHthCN{} or \HCthN{} emission from other young stellar objects\citep{Chapillon12}. These molecules exhibit bright emission signifying high gas phase abundances and column densities (N$_{\rm{Total}}$=10$^{12}$-10$^{13} \rm{cm}^{-2}$ \citep{Bergner18,Ilee21}). \CHthCN{} is an excellent probe of gas temperature due to multiple  transitions tracing a range of energy states  [lowest energy state E$_{L} \approx$5~K, spanning lower energy states over $\Delta$E$_{L}$=$\sim$100~K] that can be observed simultaneously. Each \textit{J}-transition (where \textit{J} is the rotational quanum number) has a series of \textit{K}-ladder transitions (where \textit{K} is the quantum number of angular momentum along the molecular axis) which are only sensitive to collisions, thus the ratio  between \textit{K}-transitions depends only on the gas density and temperature. The \textit{K}-transitions span a large range of temperatures and are sufficiently close in frequency to observe in one spectral setting with the \textit{Atacama Large Millimeter/submillimeter Array (ALMA)}. Analyses of rotational diagrams\citep{Goldsmith99} made from \CHthCN{} observations demonstrate an origin in  gas with a temperature between 25-50~K\citep{Ilee21}, well below the expected desorption temperature (100-124~K\citep{Corazzi21}). Thus, the brightly observed flux from this species and similar nitriles and organics like \HCthN{}, and CH$_{2}$CN\citep{Canta21} present a chemical conundrum as they should not be present in the gaseous state; rather, they should be frozen on cold grain surfaces. Where \CHthCN{} and other carbon-rich molecules reside and how they are replenished in the gas phase has a substantial impact on our understanding of prebiotic enrichment into gaseous planet atmospheres. The traditional solution to this apparent discrepancy has been to turn to grain-surface chemistry. Simple carbon- and nitrogen-bearing molecules can undergo hydrogenation or other reactions to form complex organics on the surface of a grain that then non-thermally desorb the grain intact. This solution has been used to explain the organic inventories locked onto pebbles and icy grains \citep{Oberg11ice,Bergner21}. This dust chemistry path requires both an intact photodesorption rate of \CHthCN{} from the grains of 10$^{-3}$ mols/photon\citep{Walsh14,Loomis18} and a reactive desorption efficiency\citep{Vasyunin13} of 1\%. Laboratory experiments find an intact photodesorption efficiency of \CHthCN{} to be orders of magnitude less efficient (10$^{-5}$ mols/photon\citep{Basalgete21}) and reactive desorption has not been well studied in laboratory experiments for these species. 
We posit here that there is a simpler alternative based solely on disk evolution processes that have been previously theorized and observed which link the disk gas chemistry to planet formation.  This involves two ingredients:  (1) an elevated C-to-O ratio in the gas  and (2) greater penetration of UV photons due to a reduction in the total surface area of small grain population due to pebble formation.  These combine to power a state of photo-chemistry equilibrium.

There has been mounting evidence for an elevated C-to-O gas phase ratio at large radial distances in evolved and cool gas-rich protoplanetary disks. Brighter than expected emission from small hydrocarbons such as C$_{2}$H \citep{Miotello19,Bosman21a} and complex organics such as c-C$_{3}$H$_{2}$ \citep{Cleeves21} in disks and \HCthN{} and \CHthCN{} in photon-dominated regions\citep{LeGal19} have independently suggested a C-to-O ratio well above the solar abundance ratio (C-to-O$\approx$0.55) \citep{Asplund21}. As a radical, C$_{2}$H has a short lifetime ($<$1000 years), but it is found to be abundant in the gas. Bosman et al. 2021\citep{Bosman21a} find that the only way to reproduce the high column densities observed was to increase the C-to-O ratio throughout the full disk to between 1-2. They found that it was also necessary for C$_{2}$H to exist in high density gas while remaining above the disk midplane (height/radius$\approx$0.1-0.2)
where UV photons dominate the high-energy photon budget. Together, this produces a long-lived carbon-rich gas phase cycle in photochemical equilibrium\citep{Bosman21a}.  One critical factor is that C$_2$H is predicted and is observed to exist above the disk midplane\citep{Law21emit}.  In contrast, \CHthCN{} has been observed to emit closer to the midplane\citep{Ilee21}. Extending the carbon-rich chemical environment that is used to explain C$_{2}$H observations to the midplane could allow for gas phase formation of complex molecules in the planet-forming zone.  Elevated C-to-O ratios would be necessary to supply the materials to create hydrocarbons and nitriles. However, this alone would not alleviate the issue of \CHthCN{} and \HCthN{} existing in the gas at temperatures at which they should be frozen and locked onto grains. 

A non-thermal desorption mechanism is needed to increase the number of molecules that are being desorbed from the grains, either intact or as fragments of larger molecules. If the disk atmosphere is small-dust rich, with a gas-to-dust ratio close to the typical interstellar medium (ISM) mass ratio (gas-to-dust = 100) then UV radiation cannot penetrate deep into the disk.  In this case,  only high energy radiation (X-rays and cosmic rays) can penetrate the dust and gas and drive the chemistry within and near the midplane. Small dust grains are the main opacity source for UV photons which are readily produced by the young and active star. As the protoplanetary disk evolves, small dust grains agglomerate and eventually settle downwards and drift inward radially as they grow in size\citep{Andrews18_dust,Birnstiel12}, decreasing the total surface area of the small grain population. As the main opacity source begins to deplete, UV photons can penetrate deeper into the disk. Here, complete desorption of \CHthCN{} or \HCthN{} is inhibited due to the dense and cold environment\citep{LeGal19}. This mechanism would have an effect on all molecules residing on grains, and a brief discussion regarding this can be found in the Supplemental Materials, in the section `Implications on other molecules'.

\begin{figure*}\centering
\resizebox{1\linewidth}{!}{\includegraphics{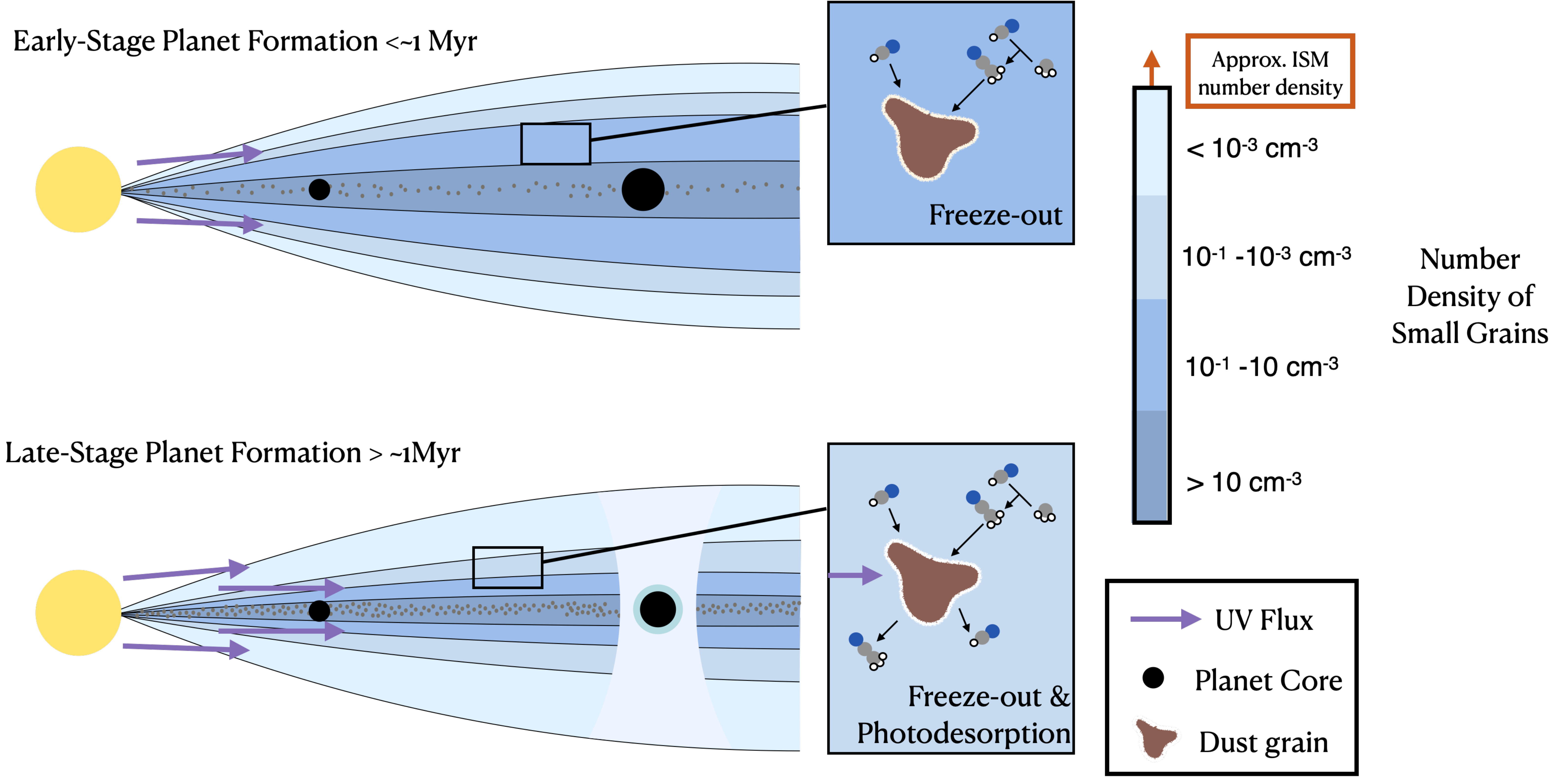}}
\caption{\textbf{A schematic highlighting the physical evolution of a disk and how that physical environment can affect the chemistry.} At the top, we show a disk with a large amount of small dust that acts to block UV photons. As the small dust settles, UV photons make their way deeper into the disk, allowing for photodesorption of complex species off grains. Now, there is a cycle of carbon chemistry that can be observed in the gas phase.}
\label{fig:schematic}
\end{figure*}

The combined effect of a high C-to-O ratio and excess UV flux allows for complex molecules to exist within the gas phase at cold, midplane temperatures. Thus bright hydrocarbon and nitrile (i.e. C$_{2}$H, CH$_{3}$CN) emission from the cold midplane acts as a signpost for an evolved dust population, coincident with an advanced stage of planet formation including the accumulation of gas-giant atmospheres. The chemical scenario we put forward, comparing early and late-stage disk environments is shown in Figure \ref{fig:schematic}. The disks observed in the Molecules with ALMA at Planet-forming Scales  (MAPS) large program\citep{Oberg21} provide plausible support for this theory. The youngest disks in the MAPS sample reside around IM Lup and AS 209 which are on the order of 1-2 Myr old\citep{Oberg21}. Towards the youngest source, IM Lup, there is no detection of \CHthCN{} nor \HCthN{}. AS 209 has detections of \CHthCN{} and \HCthN{} which suggest both molecules emit from z/r$>$0.1. The disk systems surrounding stars $>$6~Myr have \CHthCN{} detected at z/r$<$0.1 as determined by the modeled thermal structure and detected rotational temperature from the K-ladder transitions. This complex organic-rich midplane will influence the atmospheres of planetary companions. It is worth noting that the oldest disk systems in the MAPS program exist around Herbig stars while the youngest systems are T Tauri stars. The UV-bright spectrum innate to Herbig stars may additionally influence the push of complex molecules towards the planet-forming midplane. Currently, the MAPS sample contains the bulk of the resolved \CHthCN{} and \HCthN{} data towards protoplanetary disks. To explore the effect of the stellar spectrum on \CHthCN{} and \HCthN{} emission more disks with a wide range of stellar host masses would need to be observed. 

To test the validity of this proposed end-stage chemistry, we first explored single point models representative of the disk midplane with corresponding cold temperatures and high gas densities (approximately 35~K, 5$\times$10$^{11}$ mol/cm$^{-3}$ ). We varied the gas-to-dust mass ratio and initial carbon abundance among other physical and chemical variables including the nitrogen abundance, dust extinction, and ionization rate. We found that the gas phase \CHthCN{} abundance was the most sensitive to the gas-to-dust ratio and initial gas phase carbon abundance. We then produced a thermo-chemical model representing the disk around the Herbig Ae star HD 163296. This disk is old (approx. 7.6 Myr)\citep{Bergner19} and is nearby at 101 pc\citep{Gaia18} and has observed jets and winds \citep{Booth21,Xie21}. The HD 163296 disk has been widely observed in multiple gas and dust tracers with high spatial resolution ($\sim$10~au) with clear gap and ring structures typically assumed to be associated with active planet formation\citep{Andrews18, Oberg21}. There is bright \CHthCN{} emission coming from $\sim$35~K gas as well as bright emission from \HCthN{} and HCN\citep{Ilee21,Bergner21a}. Our modeling efforts follow that of Zhang et al. 2021\citep{Zhang21}, and Calahan et al. 2021b\citep{Calahan21b} which set up a thermo-chemical model of the HD 163296 disk by reproducing the mm dust-continuum observations, the spectral-energy distribution (SED), the full line intensity and morphology of six CO isotopologue transitions, and the vertical distribution of the optically thick lines. The model of HD~163296 started with a gas-to-small dust ratio of the disk equal to 500\citep{Zhang21} which was needed to reproduce the disk's SED. 

\begin{deluxetable}{c|ccc|ccc}
\label{tab:model_params}
\tablecolumns{7}
\tablewidth{0pt}
\tablecaption{Modeling Parameters: TW Hya \& HD 163296}
\tablehead{\textbf{Parameters} & & \textbf{TW Hya}&&  &\textbf{HD 163296}&\\
     &Gas& Small Dust & Large Dust & Gas & Small Dust  & Large Dust$^{a}$}
\startdata
Mass (M$_{\odot}$) & 0.025& $1.0\times 10^{-4}$ &$4.0\times 10^{-4}$ & 0.14&$2.6 \times 10^{-5}$ & 0.024\\
$\Psi$ & 1.1& 1.2 &1.2 & 1.08 &1.08&1.08\\
$\gamma$ & 0.75& 0.75 & 1.0 & 0.8&0.8&0.1\\
h$_{c}$ (au) & 42 & 42 & 8.4 & 8.44&8.44& n/a\\
r$_{c}$ (au) & 400 & 400 & 400 & 165&165& n/a\\
r$_{in}$ (au) & 0.1 & 0.5& 1 & 0.45 & 0.45&0.45\\
r$_{out}$ (au) & 200 &200&200& 600& 600&240\\
\enddata
\tablecomments{Final values of the TW Hya and HD 163296 models that reproduce CO, HD, \CHthCN{}, HCN, and \HCthN{} observations when available. r$_{in}$ and r$_{out}$ are the radial inner and outer limits of the disk, beyond these limits there is assumed to be no gas nor dust. 
$^{a}$ The surface density of the large dust distribution in HD 163296 is empirically set by continuum observations, thus it is not smooth and it is not dictated by the parametric equations.}
\end{deluxetable}

\section*{\large Results} 

\begin{figure*}\centering
\resizebox{1\linewidth}{!}{\includegraphics[scale=.4]{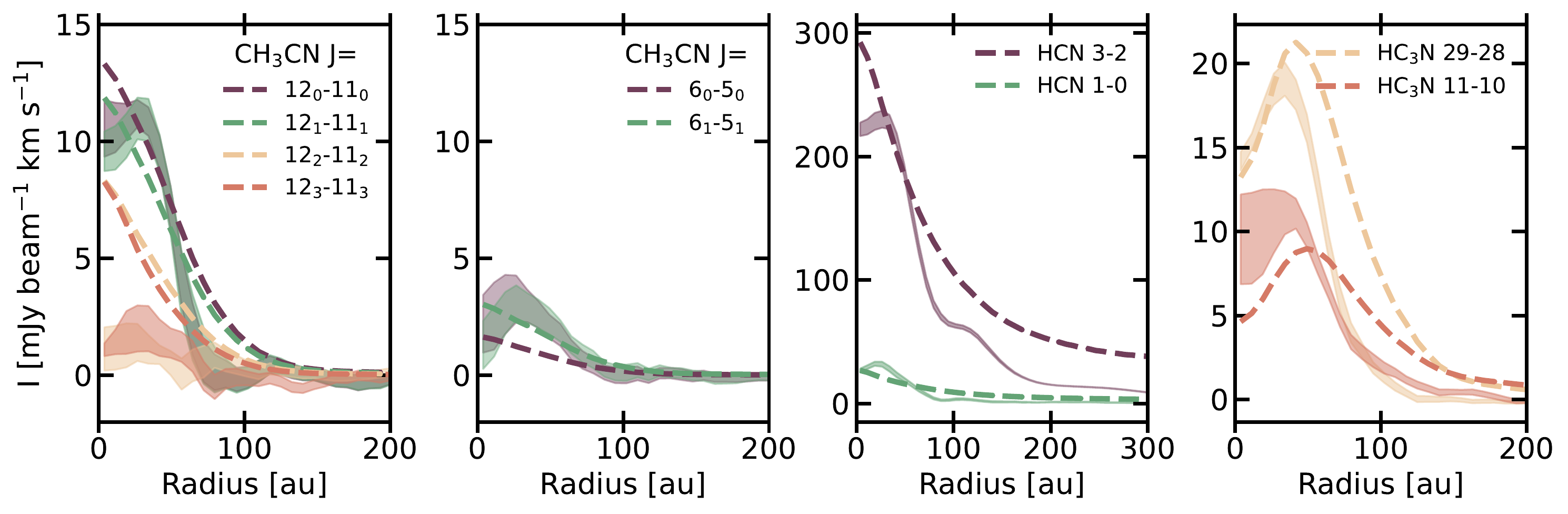}}
\caption{\textbf{ Radial intensity profiles of the observed complex organic molecules and HCN towards the disk around HD 163296.} The molecules shown are \CHthCN{}, HCN, and \HCthN{} (left to right with two \textit{J} transition of \CHthCN{} shown in the first two panels). Solid thick lines in the background correspond to the observations derived from Ilee et al 2021\citep{Ilee21} and Guzm\'an et al. 2021\citep{Guzman21} (which utilized Law et al. 2021\citep{Law21rad}) while dashed lines correspond to modeled radial profiles. Our final model includes an increase in C/O ratio beyond 20~au and a depletion of small dust and represents 1~Myr of chemistry. This model can simultaneously fit the flux, line ratios, and general morphology of the observed radial profiles of these complex organic molecules.}
\label{fig:observations}
\end{figure*}

To produce this end-stage chemical environment, we deplete the small dust mass by a factor of 10 throughout the disk of HD 163296 making the new gas-to-dust mass ratio above the pebble disk midplane equal to 5,000. We then enhanced the initial gas phase carbon abundance in the system in the form of C, CH$_{4}$, or C$_{2}$H, increasing the overall gas phase carbon-to-oxygen ratio in both disks to above unity. There are two possible sources for excess carbon in protoplanetary disks, either through the destruction of refractory carbon grains\citep{Bosman_refrac} or CO depletion through mechanisms such as reactions with ionized molecules and atoms such as He$^{+}$ or H$_{3}^{+}$\citep{Schwarz18_1} and a series of chemical and freeze-out processes. We note that while signatures of CO depletion are found in the HD 163296 disk\citep{Zhang21,Calahan21b}, CO destruction likely only occurs in environments with a high cosmic-ray flux [2$\times$10$^{-17}$s$^{-1}$]\citep{Schwarz18_1}. Additionally, CO destruction would supply equal amounts of carbon and oxygen, while we seek to enhance carbon over oxygen.  Our thermo-chemical model is run for a full megayear (Myr), after which the chemistry reaches an equilibrium. Notable chemical feedback due to dust evolution is predicted to occur over scales of $\sim$1~Myr\citep{Krijt18,VanClepper22}, thus we use $\sim$1~Myr as an approximate length of time for the chemical environment to transition from `early-stage' to `late-stage' dust evolution and subsequently exist in a state of photochemical equilibrium. Gas phase chemical reactions and rates were taken from the chemical network derived in Bosman et al. 2018\citep{Bosman18} which in turn relies on the UMIST Database for Astrochemistry Rate12 version\citep{McElroy13} (see Methods section for more details). We found that regardless of the carrier of carbon, using a C/O = 1-2 and a factor of 10 depletion of small dust roughly reproduces line ratios and fluxes of the radial intensity profiles of \CHthCN{}, \HCthN{}, and HCN (see Figure \ref{fig:observations}). In this proposed scenario, $\sim$96\% of the total mass of \CHthCN{} continues to reside frozen out onto grains, but the increase in UV flux allows for sufficient \CHthCN{} to exist and be formed in the gas to reproduce observed radial intensity profiles and column densities.

\begin{figure*}\centering
\resizebox{1\linewidth}{!}{\includegraphics[scale=.4]{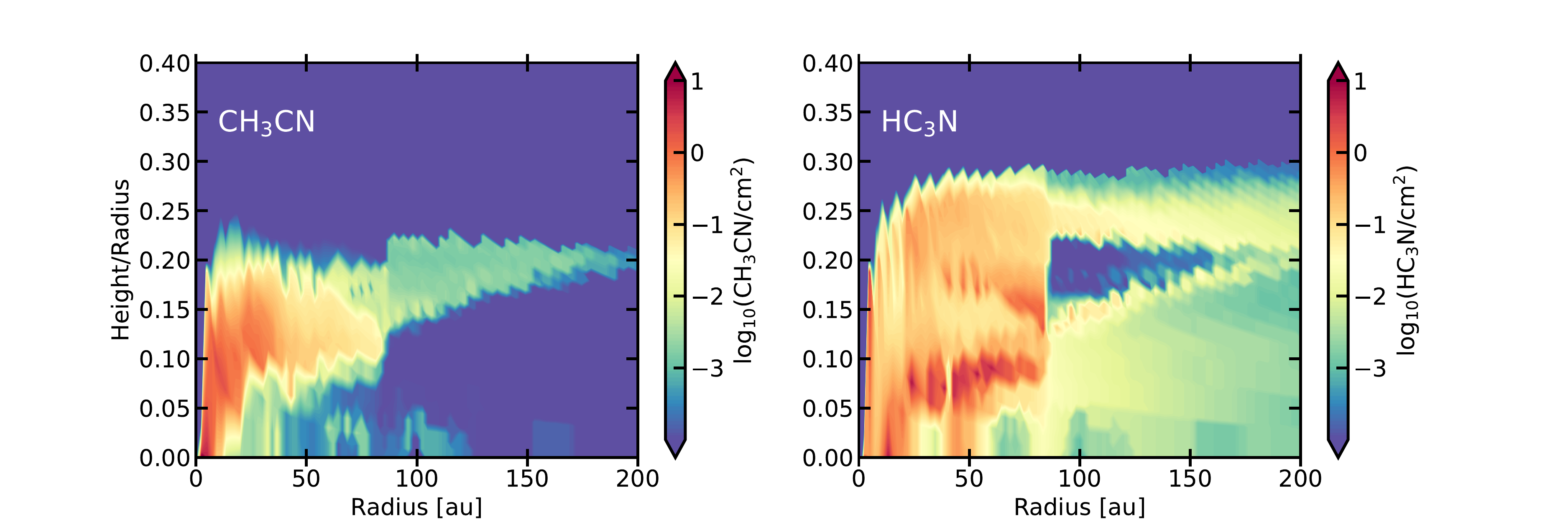}}
\caption{\textbf{ The radial and vertical number density distributions of \CHthCN{} and \HCthN{}} The density distributions were determined by our final HD 163296 model with an increase in C/O ratio beyond 20~au and a depletion of small dust. This distribution produced the radial profiles shown in Figure \ref{fig:observations}. A z/r$<$0.1 is good approximation for the midplane of the disk, and both \CHthCN{} and \HCthN{} emit partially from the midplane. A comparison can be made with Figure \ref{fig:2d_abundances_noUV} which lacks the inclusion of small dust depletion.}
\label{fig:2d_abundances}
\end{figure*}

Figure \ref{fig:2d_abundances} shows the two-dimensional number densities of \CHthCN{} and \HCthN{}. The inner $\sim$20 au exhibited brighter or more centrally peaked emission than was seen in observations if the C-to-O ratio = 2 throughout the full disk. To counteract this, we set the C-to-O ratio~=~0.47, or the ISM ratio, inside of ($\sim$20~au) where Zhang et al. 2021\citep{Zhang21} found an ISM ratio of \Htwo{}/CO (See Supplementary Figure 2). The inner disk emission remains slightly brighter than observed in HCN and \CHthCN{} (see Figure \ref{fig:observations}) with this alteration. This could be accounted for via a depletion in the nitrogen abundance within the N$_{2}$ ice line, a strong buildup of pebbles around the water ice line ($\sim$5au), a lower HCN desorption energy, or a combination of these effects. Additionally, the observations may also be affected by processes such as beam smearing and a higher dust opacity than modeled. A comparison model without a depletion in small dust mass is shown in Supplementary Figure 5.

\section*{\large Discussion}

\HCthN{} and \CHthCN{} are built up in the gas from simple carbon and nitrogen-based volatiles (see Methods section ``Chemical Reactions" for details). A reservoir of these more simple molecules is maintained in the gas phase due to photodesorption from dust grain surfaces by an enhanced UV-field. The main destruction products from \CHthCN{} and \HCthN{} are simple volatiles that can cycle back to create larger nitrile molecules including their original parent molecule.  While there are enough complex organic molecules in the gas phase to observe bright emission, the majority of the \CHthCN{}, \HCthN{}, and HCN near the midplane still remains frozen out onto grains. The carbon-rich gas reservoir and UV-dominated disk together allow for a cycle of carbon chemistry to remain active in the gas phase. This implies that actively accreting gas giants will build their atmosphere out of this complex nitrile-enhanced material. 
Gas giant planets with atmospheres containing a chemical make-up with a C-to-O ratio $>$1 can be explained by the natural evolution of dust and the observed high C-to-O ratios within protoplanetary disks. This environment could be extended to the inner disk in some cases, as studies such as Najita et al. 2011\citep{Najita11} and Anderson et al. 2021\citep{Anderson21} posit a higher than solar C-to-O ratio within the inner disk to account for Spitzer observations of HCN and C$_{2}$H$_{2}$ around T~Tauri stars. 

The implications of this end-stage chemistry are far-reaching. Primarily, we put forward a new chemically and physically coupled picture of planet formation and its direct impact on the chemistry of forming planet atmospheres. Within the early stages of planet formation, pebble growth and accretion into solid planet cores has begun, and chemistry onto these cores are influenced primarily by the chemical make-up of large pebbles and their icy mantles.  The chemistry active in the midplane during this stage of formation is dominated by X-rays and cosmic rays\citep{Woitke09}.  After timescales of order 1 Myr,  at the end-stage of pebble formation, small grains have grown and settled towards the midplane, allowing for UV photons to penetrate deeper in the disk than previously assumed. This excess UV flux in concert with an above solar C-to-O ratio produces a carbon-rich gas phase cycle of production and destruction in the gas surrounding actively forming planets. Bright emission from \CHthCN{}, \HCthN{} and other carbon-rich molecules emitting at cold temperatures are a sign-post of this evolved chemical environment. These forming gas giants may accrete this surrounding gas and its chemical signature into their atmosphere, and may be responsible for high C-to-O measurements that have been observed in exo-planetary atmospheres as compared to their host star\citep{Brewer17}.
Secondly, the deeper penetration of UV photons pushes the CO and H$_{2}$ self-shielding layers deeper into the disk. Due to this, there are more ions such as C$^{+}$ and H$^{+}$ throughout the disk than would have been present with a regular (ISM) level of small dust surface area. As the ion-fraction increases so does the area of the disk that is subject to magneto-rotational instability (MRI). In our final model, the UV field increases by a factor of 3-5 in the atmosphere and between 1-2 near the midplane, and thirteen times more gas mass is MRI-active, most notably the gas midplane within 10~au, and then at a z/r=0.1-0.2 within 30 au (see Figure 4). This could be a source of accretion\citep{Gammie96}  for older disks. Our proposed UV-dominated carbon-rich gas phase chemistry is a major shift in our understanding of the astrochemistry of planet formation. This transition will have a strong effect on the composition of actively forming planets, disk MRI-activity, and subsequent disk accretion. 

\begin{figure*}\centering
\resizebox{1\linewidth}{!}{\includegraphics{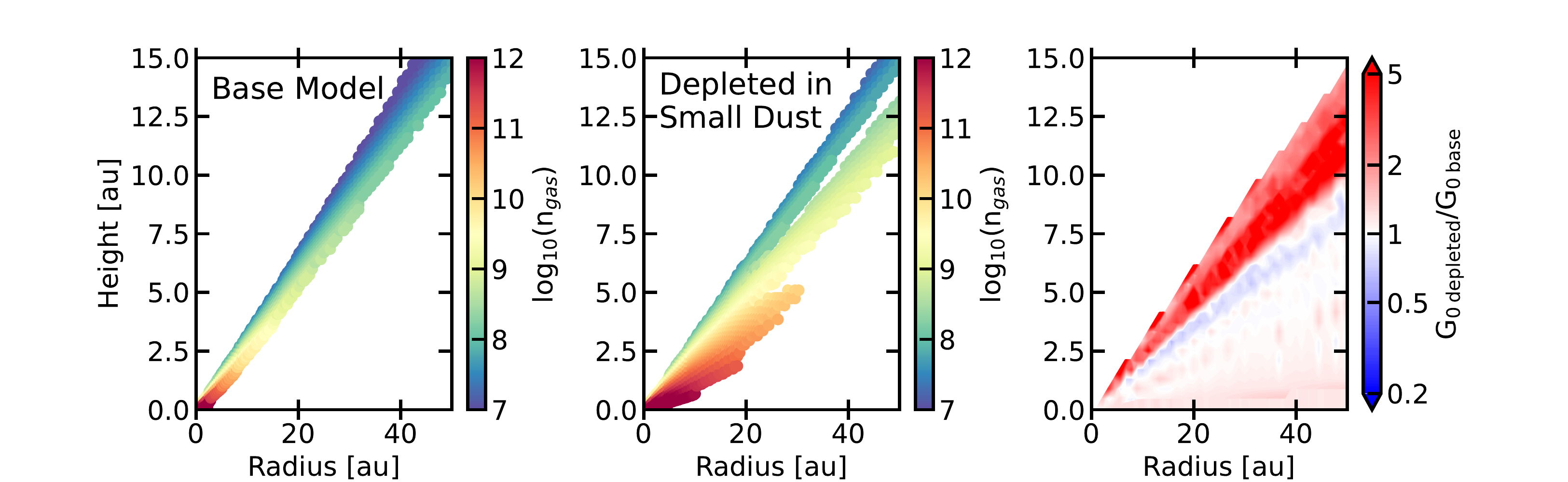}}
\caption{\textbf{A comparison of the MRI active region within a non-elevated UV environment and an elevated UV environment.} The two left-most plots show regions within the HD 163296 disk which are MRI activated in an environment where small dust is depleted by a factor of ~10 in total mass (gas/dust = 5,000 middle plot) and where small dust is not depleted (gas/dust=500 left). The color corresponds to the gas density in mol/cm$^{3}$, highlighting the substantial increase in mass that is MRI-active. The right-most plot shows a comparison of the UV field strength through a `regular' disk and a dust depleted disk. The UV field increases upwards of a factor of 3-5 in the atmosphere of the disk and 1-2 below a z/r=0.2. This has a notable effects on the chemistry of the midplane.}
\label{fig:MRI}
\end{figure*}

\section*{\large Methods}
\subsection*{Observations}
For a thorough description of the observations and the techniques used to obtain the CLEAN-ed images and radial profiles of \CHthCN{}, HCN, and \HCthN{} towards HD 163296 see Czekala et al. 2021\citep{Czekala21}, Law et al. 2021\citep{Law21rad}, and \"Oberg et al. 2021\citep{Oberg21}. A description of the observations of \CHthCN{} towards TW Hya can be found in Loomis et al. 2018a\citep{Loomis18}. A brief summary of these observations are as follows.

\subsubsection*{- HD 163296 -}

The data for HD 163296 is from the MAPS large program (Project ID 2018.1.01055.L)\citep{Oberg21}. The image cubes were produced using the \texttt{tclean} task in the Common Astronomy Software Applications (CASA) package, version 6.1.0\citep{McMullin07}. Keplerian masks based on the disk geometric parameters were used in the CLEANing process and the final images were then corrected for the Jorsater \& van Moorsel effect to ensure that the image residuals are in units consistent with that of the CLEAN model. For all lines, we used the beam-circularized and uv-tapered images, which had synthesised beam sizes of 0."3. All radial intensity profiles were generated by deprojecting and azimuthally-averaging zeroth moment maps using the GoFish python package.

\subsubsection*{-TW Hya-}

\CHthCN{} was observed towards TW Hya as a part of ALMA project 2016.1.01046.S. Each emission line was individually imaged using \texttt{CLEAN} and the synthesised beam for each transition were matched using small uv-tapers to 1$.\!\!^{\prime\prime}$05 x 0$.\!\!^{\prime\prime}$83. A Keplerian mask was used to extract the flux for each transition. See Loomis et al. 2018a\citep{Loomis18} for more details.

\subsection*{Modeling}

Simple single point models of the disk environment were utilized to quickly and efficiently understand the chemical impacts of different physical and initial chemical conditions. Our chemical network is derived from Bosman et al. 2018\citep{Bosman18} which in turn is derived from the UMIST ``RATE12'' network\citep{McElroy13}, and for this study we disregarded most grain-surface chemical reactions in order to isolate gas-phase formation of \CHthCN{}. After our results from the modeling suggested a higher gas/dust ratio and carbon content, we turned to more comprehensive thermo-chemical codes which model the thermal physics and chemical evolution throughout the whole disk. The code RAC2D (https://github.com/fjdu/rac-2d)\citep{Du14}, was used to create models of the disks around TW Hya and HD 163296. The 2D temperature, density, and molecular abundance results were used to simulate observations of each disk with a raytracing code: RADMC-3D\citep{radmc3d}. A brief description of the physical code of RAC2D is given below; a detailed description of the code can be found in Calahan et al 2021a\citep{Calahan21a}.

RAC2D takes into account a gas and dust structure and stellar radiation field and computes the gas and dust temperature and chemical structure over time. Our model consists of three mass components: gas, small dust, and large dust grains. The spatial extent of each component is given by a global surface density distribution \citep{Lynden-Bell74}, which is widely used in protoplanetary disk modeling and corresponds to the self-similar solution of a viscously evolved disk.

\begin{equation}
\Sigma(r)=\Sigma_{c}\left(\frac{r}{r_{c}}\right)^{-\gamma}\exp{\left[-\left(\frac{r}{r_{c}}\right)^{2-\gamma}\right]},
\label{surfden_equn}
\end{equation}

\noindent \(r_{c}\) is the characteristic radius at which the surface density is \(\Sigma_{c}/\rm{e}\) where $\Sigma_{c}$ is the characteristic surface density,  and \(\gamma\) is the power-law index that describes the radial behavior of the surface density.

A 2D density profile for the gas and dust populations can be derived from the surface density profile and a scale height:

\begin{equation}
    \rho(r,z) = \frac{\Sigma(r)}{\sqrt{2\pi}h(r)} \exp{\left[-\frac{1}{2}\left(\frac{z}{h(r)}\right)^{2}\right]},
\label{density_equn}
\end{equation}

\begin{equation}    
    h=h_{c}\left( \frac{r}{r_{c}}\right)^{\Psi} ,
\label{heigh_equn}
\end{equation}

\noindent where \(h_{c}\) is the scale height at the characteristic radius, and \(\Psi\) is a power index that characterizes the flaring of the disk structure. The modeling parameters used for TW Hya and HD 163296 are shown in Table 1.

For both TW Hya and HD 163296 models, each dust population follows an Mathis, Rumpl, and Nordsieck (MRN) grain distribution \(n(a) \propto a^{-3.5}\) \citep{Mathis77}, where `a' indicates the size of the grain. The small dust grains have radii between $5 \times 10^{-3}$ - $1 \mu$m, and the large grains have radii between $5 \times 10^{-3}$ - $10^{3} \mu$m.  The large dust population is settled in the midplane with a smaller vertical extent and radial extent (gas extends $\sim$5 and 2.5 times above and beyond, respectively). This settled large grain population is the result of dust evolution, namely growth in concert with vertical settling to the midplane and radial drift. For the HD 163296 model, the large grain population has a unique, non-smooth, surface density profile that reproduces the millimeter continuum observations of the HD 163296 disk\citep{Isella18,Zhang21}.  Opacity values for the dust are calculated based on Birnstiel et al. 2018\citep{Birnstiel18}. Large dust grains consist of water ice\citep{Warren08}, silicates\citep{Draine03}, troilites and refractory organics\citep{Henning96}. Small dust grains consist of 50\% silicates and 50\% refractory organics. Further discussion on the modeling efforts for TW Hya can be found in Calahan et al. 2021a\citep{Calahan21a} and for HD 163296 in Calahan et al. 2021b\citep{Calahan21b}.

The thermo-chemical model resulted in 2D distributions of the gas and dust temperature, density, and molecular abundances. These were used as inputs for the raytracing code RADMC-3D\citep{radmc3d}. The molecular properties of \CHthCN{}, \HCthN{}, and HCN were taken from the Leiden Atomic and Molecular Database (LAMDA\citep{Schoier05}), with some molecular parameters updated according to data from the Cologne Database for Molecular Spectroscopy (CDMS\citep{Muller01,Muller05}). The result from RADMC-3D was a 3D image cube of the molecular emission across velocity space. We utilized \texttt{GoFish}\citep{GoFish} to compress these 3D images into zeroth moment maps and radial profiles which were then directly compared to the azimuthally-averaged observations.

\subsection*{Chemical Reactions}
The chemical reaction network utilized in this work comes from Bosman et al. 2018\citep{Bosman18} which in turn relies on the Rate12 version of the UMIST Database for Astrochemistry\citep{McElroy13}: a network that is widely used across astrochemical modeling efforts. In our network there are 6,302 gas phase reactions and we have limited the grain-surface reactions to twelve. We kept grain-surface reactions that involved the formation of H$_{2}$, CH$_{2}$OH, CH$_{3}$OH, CO$_{2}$, H$_{2}$O and NH$_{3}$ due to each of these species being well-studied in the laboratory and come with strong evidence for active two-body chemistry on dust grains\citep{Allen77,Hasegawa92,Goumans08,Ioppolo08,Ioppolo10,Fuchs09,Oba12,Ruffle01} (see Table 2). Both thermal adsorption and desorption are taken into account for every molecule in the network. Our network contains molecules with at most eleven carbon atoms, and at most twelve total atoms. We run our thermo-chemical model for 1~Myr as the \CHthCN{} and \HCthN{} chemistry reaches an equilibrium after this time period (see Supplementary Figure 3).  In our evolved disk model, the main formation pathways for \CHthCN{} once the chemistry has reached equilibrium is as follows:

\[\rm{CH_{3}^{+} + HCN \rightarrow CH_{3}CNH^{+} + e^{-} \rightarrow CH_{3}CN + H}\] 

\noindent This reaction in the gas phase is the primary formation pathway for \CHthCN{}, thus there is a strong reliance on HCN existing in the gas phase even at cold temperatures below its measured sublimation temperature ($\sim$85-103~K\citep{Bergner22}) and an ionization source (UV photons) to produce CH$_{3}^{+}$. \HCthN{} can be readily produced by a number of different ways including: 

\[\rm{N + CH_{2}CCH \rightarrow HC_{3}N + H_{2}}\]
\[\rm{CN + C_{2}H_{2} \rightarrow HC_{3}N + H}\]
\[\rm{H + C_{3}N^{-} \rightarrow HC_{3}N + e^{-}}\]

\noindent The formation of \HCthN{} strongly relies on the existence of carbon-rich molecules including radicals (CN) and ions (C$_{3}$N$^{-}$).

\begin{deluxetable}{c|c}
\label{tab:dust_react}
\tablecolumns{7}
\tablewidth{0pt}
\tablecaption{Dust Surface Reactions}
\tablehead{\textbf{Reaction} & \textbf{Reference}}
\startdata
gH + gH $\rightarrow$ gH$_{2}$ & Hasegawa, Herbst, \& Leung 1992\citep{Hasegawa92} \\
gH + gOH $\rightarrow$ gH$_{2}$O & Ioppolo et al. 2010\citep{Ioppolo10} \\
gH + gH$_{2}$O$_{2}$ $\rightarrow$ gH$_{2}$O + gOH & Ioppolo et al. 2008\citep{Ioppolo08}\\
gH + gCH$_{3}$OH $\rightarrow$ gH$_{2}$ + gCH$_{2}$OH & Extrapolated from Fuchs et al. 2009\citep{Fuchs09}  \\
gH + gCH$_{2}$OH $\rightarrow$ gCH$_{3}$OH & Extrapolated from Fuchs et al. 2009\citep{Fuchs09}  \\
gH$_{2}$ + gOH $\rightarrow$ gH$_{2}$O + gH & Oba et al 2012\citep{Oba12} \\
gOH + gCO $\rightarrow$ gCO$_{2}$ + gH $^{a}$ & Ruffle \& Herbst 2001\citep{Ruffle01}\\
gO + gCO $\rightarrow$ gCO$_{2}$ $^{a}$& Goumans \& Brown 2008\citep{Goumans08} \\
gO + gHCO $\rightarrow$ gCO$_{2}$ + gH & Goumans \& Brown 2008\citep{Goumans08} \\
gH + gNH$_{2}$ $\rightarrow$ gNH$_{3}$ & Allen \& Robinson 1977\citep{Allen77}\\
\enddata
\tablecomments{The complete list dust surface reactions accounted for in this study.
$^{a}$Have additional special treatment for three body reactions}
\end{deluxetable}

\section*{\large Supplementary Information}

\subsection*{TW Hya}
\begin{figure*}\centering
\resizebox{1.0\linewidth}{!}{\includegraphics{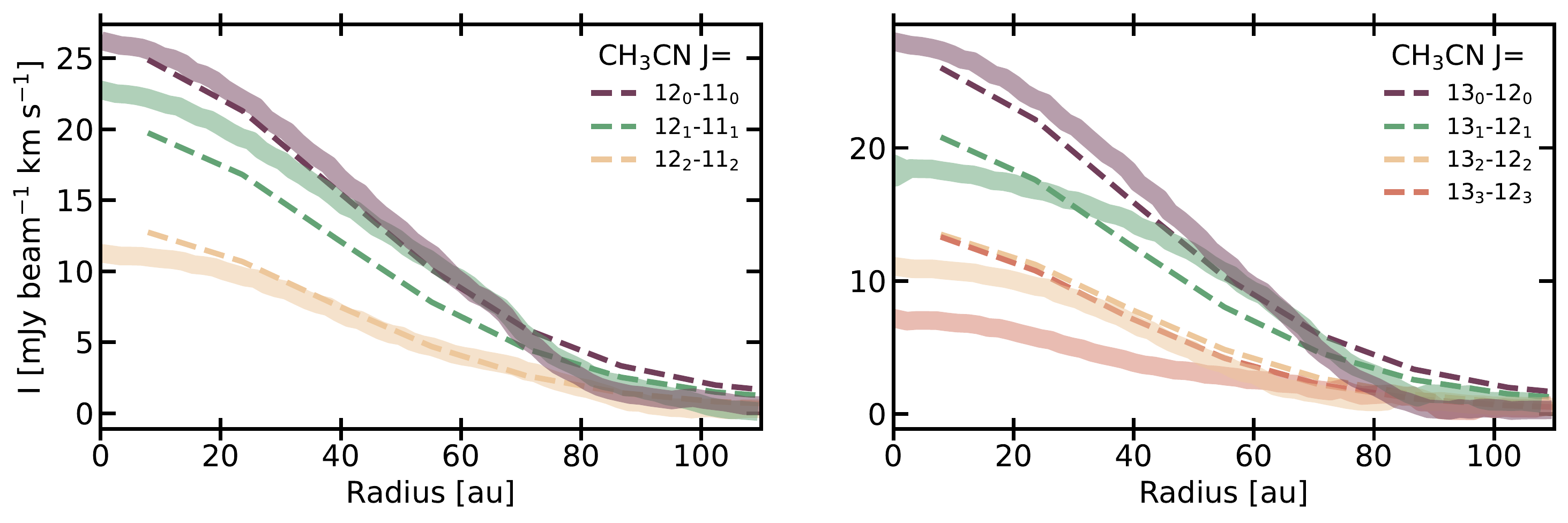}}
\caption{  A comparison of observed CH$_{3}$CN\citep{Loomis18} towards TW Hya and a thermo-chemical model. The model is run for 1 Myr with a gas-to-dust ratio equal to 250 and a C-to-O ratio equal to 1.0. Observed emission is shown as the solid lines while modeled results are the dashed lines. The left and right panels show different J$_{K} $ transitions.}
\label{fig:TWHya}
\end{figure*}

We additionally produced a thermo-chemical model representing the disk around T Tauri star TW Hya. TW Hya is approximately 10 Myrs old\citep{Thi10} and hosts the closest Class II disk at 59.9 pc\citep{Gaia18}. TW Hya has been widely observed in multiple gas and dust tracers with high spatial resolution ($\sim$10~au)\citep{Andrews12,Huang18}. It also exhibits bright \CHthCN{} coming from $\sim$33~K gas\citep{Loomis18}. Our modeling efforts of TW Hya follow that of Calahan et al. 2021a\citep{Calahan21a} which sets up a thermo-chemical model of the disk by reproducing the spectral-energy distribution (SED), the full line intensity and morphology of seven CO isotopologue transitions, and an HD J=1-0 observation from the \textit{Hershel} PACS instrument\citep{Bergin13}. To reproduce all CO radial profiles as well as the HD flux, the small dust in the upper layers of the atmosphere of the TW Hya disk were effectively depleted slightly\citep{Calahan21a} due to having a slightly lower flaring angle than the gas population. This was enough small dust depletion to reproduce the available \CHthCN{} lines from Loomis et al. 2018a given a C/O ratio equal to 1.0 (see Figure \ref{fig:TWHya}). This disk is an additional piece of evidence supporting our evolved chemistry proposal.

\subsection*{Implications on Other Molecules}

This introduction of a late-stage photo-chemical equilibrium was motivated by observations of \CHthCN{}, \HCthN{}, and HCN. However, the late-stage chemistry would have an effect on other molecules as well. We predict in addition to these organic molecules, molecules such as C$_{X}$H, C$_{X}$H$_{2}$, and HC$_{X}$N will be abundant in the gas, enhanced by the cycle of carbon chemistry that reproduces observed \CHthCN{} and \HCthN{}. The main carbon carrier, CO, is largely unaffected by the increase in photo-chemistry. CO is largely in the gas phase in previous models that do not include this `late-stage' chemistry and in the region in which \CHthCN{} was added into the gas, CO was already primarily in the gas phase. We find a slight depletion of CO in the upper atmosphere of the disk due to the CO photodissociation layer being pushed down, and there is a slight enhancement of CO in the midplane due to the UV-enhancement. However this accounts for less than 1\% of the total abundance of CO thus did not have a strong impact on the modeled radial profiles. Another key molecule is H$_{2}$O. In our models, we do not initialize our model with H$_{2}$O, thus there is very little water to be affected by this `late-stage' chemistry. Observational results of Du et al. 2017\citep{Du2017} support this as they found a low overall abundance in H$_{2}$O in disks with a survey of 13 protoplanetary disks. \CHthCN{} nor \HCthN{} would be seen to be at a high abundance if gas-phase H$_{2}$O was in high abundance in the disk as it would disrupt the carbon-rich chemistry.

\subsection*{SED Degeneracy}

The depletion of small grains will affect the observed spectral energy distribution (SED) from each disk. TW Hya’s dust population was not altered from that of Calahan et al. 2021a\citep{Calahan21a} and continues to match its observed SED. A depletion of a factor of 10 in the small dust population around HD 163296 would cause a dimming in the mid-infrared part of the SED. However, the small dust abundance, the distribution of dust grain size, and the assumed dust opacity sources are degenerate in the ways they may affect the disk SED. We modeled protoplanetary disk SEDs using the code TORUS\citep{Harries04,Harries19}. TORUS is a Monte Carlo radiative transfer code utilizing radiative equilibrium\citep{Lucy99} and silicate grains\citep{Draine84}. In Figure \ref{fig:SED_example}, we show a series of SEDs produced from models motivated by our thermo-chemical model of HD 163296 including the stellar parameters and dust distribution  in Table 1 from Calahan et al. 2021b\citep{Calahan21b}. We find that by varying the minimum dust grain radius or the power law index, we can account for a factor of 10 in UV attenuation. The total population mass, grain size, and how the grain sizes are distributed are strongly degenerate and can result in uncertainties of the dust mass by a factor of at least 10. Thus our depletion of small dust continues to reproduce all previous observables including the SED.

\subsection*{MRI Instability}
With the increase in UV flux deeper into the disk, more ions are created. Ions can be coupled with the magnetic fields that thread through the disk and interact with the bulk gas creating turbulence. This mechanism is called the magneto-rotational instability or MRI\citep{Balbus91}. Magneto-hydrodynamical processes are assumed to be present throughout protoplanetary disks due to young star’s active magnetic field, and are thought to be one of the main drivers of angular momentum transport \citep{Chandrasekhar60}. An MRI-active zone may drive the bulk of the mass transportation in the disk and activate accretion onto the star, two vital processes that determine the future of a young solar system. It is thought that planet formation may be aided within ‘dead-zones’ where MRI is non-active\citep{Gressel12}. The magnetic Reynolds number ($Re$) and ambipolar diffusion term ($Am$) are two quantities that help quantify the presence of an MRI-active zone. The Reynolds number quantifies the level of coupling between ionized gas and magnetic fields and is defined as

\begin{equation}
    Re \equiv \frac{c_{s}h}{D} \approx 1 \left(\frac{\chi_{e}}{10^{-13}}\right) \left(\frac{T}{100~\rm{K}}\right)^{1/2} \left(\frac{a}{\rm{AU}}\right)^{3/2} ,
\end{equation}

\noindent where $c_{s}$ is the sound speed, $h$ is the scale height of the disk, D is the magnetic diffusivity parameter, $\chi_{e}$ is the electron abundance, $T$ temperature and $a$ radial location in the disk \citep{Perez-Becker11}. The ambipolar diffusion term describes the coupling of ionized molecules and their interaction with neutral gas particles:

\begin{equation}
    Am \equiv \frac{\chi_{i} n_{\rm{H}_{2}} \beta_{\rm{in}}}{\Omega} \approx 1 \left(\frac{\chi_{i}}{10^{-8}}\right) \left(\frac{n_{\rm{H}_{2}}}{10^{10}\rm{cm^{-3}}}\right) \left(\frac{a}{\rm{AU}}\right)^{3/2} ,
\end{equation}

\noindent where $\chi_{i}$ is the ion abundance, $n_{\rm{H}_{2}}$ is the number density of H$_{2}$ atoms, $\Omega$ is the dynamical time, $\beta_{\rm{in}}$ is the collisional rate coefficient for singly charged species to share momentum with neutral species, and $a$ is the radial location in the disk \citep{Draine83, Perez-Becker11}. For MRI to act as a turbulent driver of neutral gas, both $Re$ and $Am$ must be sufficiently high. Simulations show that values between 0.1-100 for $Am$ can trigger significant coupling between ions and neutrals \citep{Hawley98,Bai11}. Models by Flock et al. 2012\citep{Flock12} suggest $Re \approx$ 3,300-5,000 is required to sustain sufficient turbulence with a critical $Re$ = 3,000. In our work, we assume a combined $Am >$100 and $Re >$  3,000 to signify an MRI-active zone.

We calculate the regions in which the MRI is active in the HD 163296 disk in our models with an early-stage physical environment and a late-stage environment (signified by a depletion in the small dust population mass) in Figure \ref{fig:AmRe}. Our solution allows for an increase of UV flux in the atmosphere by a factor of 3-5. More of the disk becomes ion-rich due to the CO and H$_{2}$ self-shielding layers being located deeper into the disk. The increase in the ion and electron abundance ($\chi_{i}$, $\chi_{e}$) are the key factors that enhance the Re and Am values and thus produce additional MRI activity (see Figure \ref{fig:AmRe}). As a result, thirteen times more mass within the disk becomes MRI-active including within the midplane. In the base model representing early stages of disk chemistry, the MRI activity at the midplane extends to 4~au. By depleting the small dust population by an order of magnitude, the MRI activity at the midplane then extends out to $\sim$10~au. This increase in MRI activity can contribute to the reason behind why older disk systems, such as TW Hya, are actively accreting. Meridional flows have been identified in the HD 163296 disk located at two of the largest dust gaps  (approx. 45 and 86 AU)\citep{Teague19_flows} and these vertical flows are coincident with the lower vertical limit of the MRI-active zone. The MRI activation could have an effect on the height or continue to drive meridional flows. Turbulence measurements of the HD 163296 disk have been derived in Flaherty et al. 2015 and 2017\citep{Flaherty15,Flaherty17} using molecular tracers, and they find turbulent velocities to be 5\% or less of the sound speed between 30-300~au and at all measured heights. It is not yet clear whether our MRI prediction is in tension with the observational evidence of low turbulence in HD 163296, more work needs to be done on the modeling of how MRI-active regions drive turbulence and more observational constraints are needed to constrain the inner 40~au where we find the strongest MRI-activity.

\subsection*{Supplementary Figures}

\begin{figure*}
\resizebox{1.0\linewidth}{!}{\includegraphics{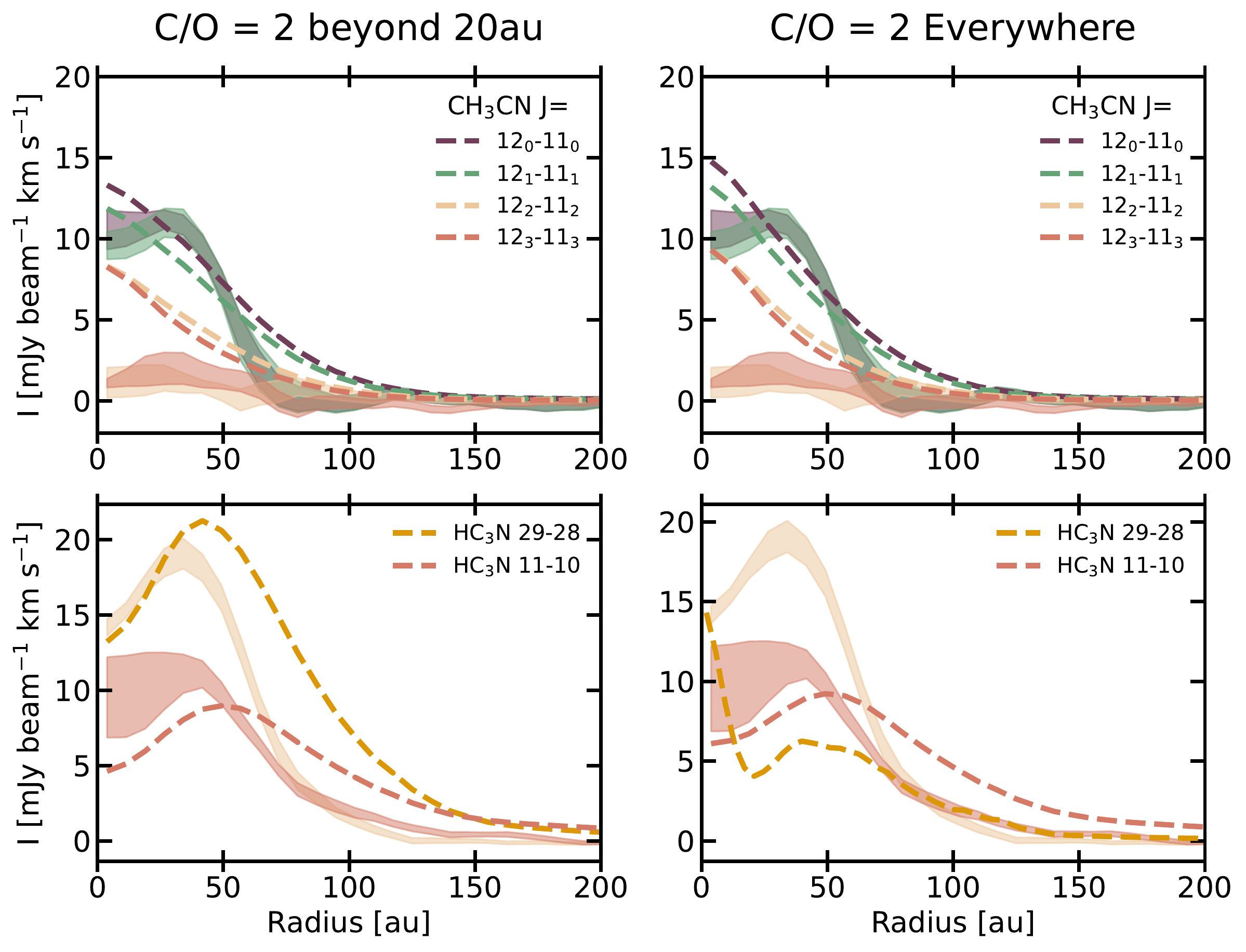}}
\caption{{ A comparison of two HD 163296 disk models with varying C-to-O ratios. The final model has a C-to-O = 0.47 within 20~au and C-to-O = 2 beyond 20~au (left). The comparison model has a C-to-O ratio equal 2 throughout the entire disk.  Dashed lines are modeled radial intensity profiles while the thick line in the background are observations, with the thickness corresponding to the uncertainty of the flux. In our final model, the C-to-O ratio is equal to what is found in the ISM, 0.47.}}
\label{fig:C-to-O_compare}
\end{figure*}

\begin{figure*}\centering
\resizebox{1.0\linewidth}{!}{\includegraphics{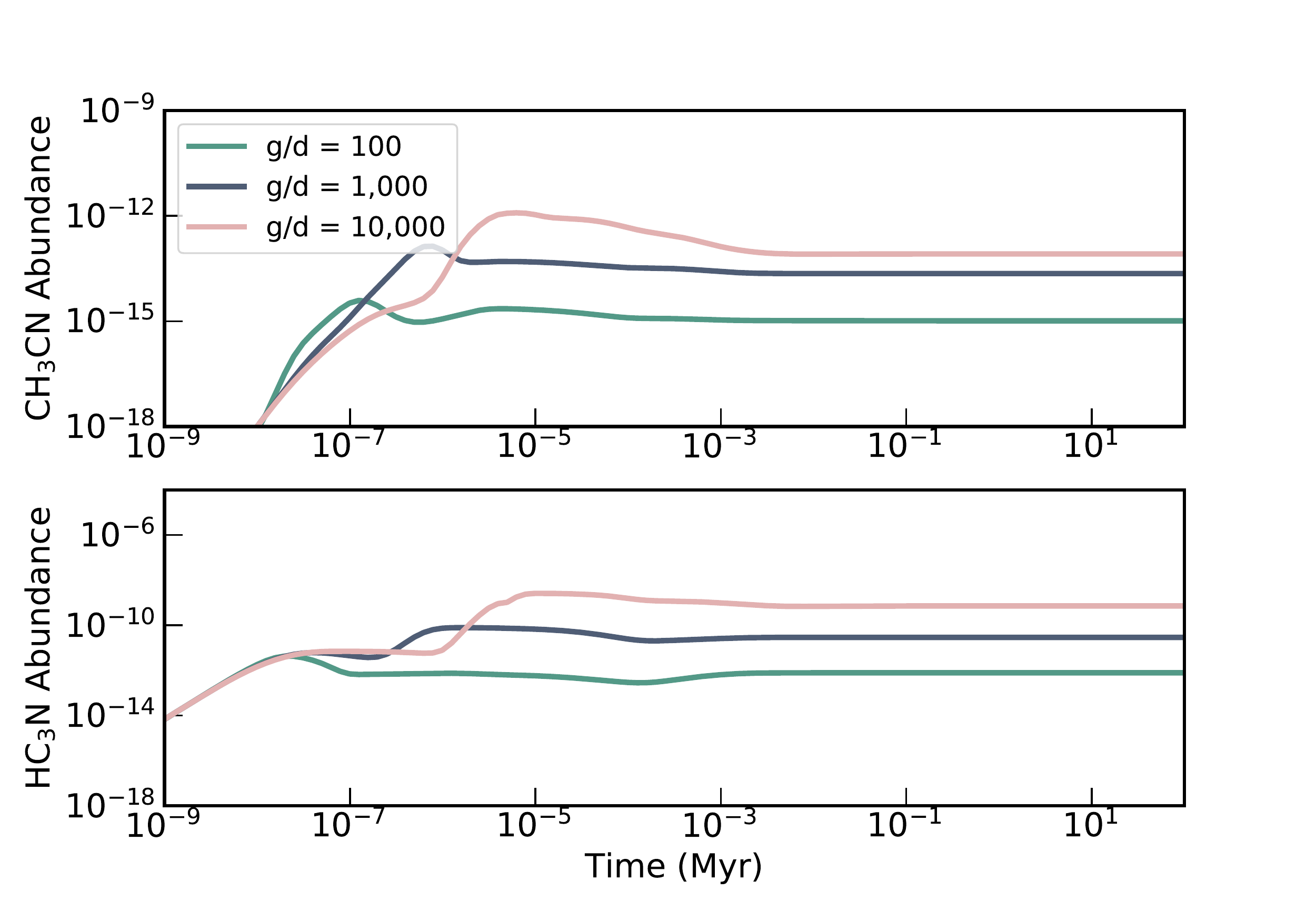}}
\caption{{The evolution of \CHthCN{} and \HCthN{} abundances relative to H$_{2}$ given different gas to dust ratios. This considers an environment with a 35~K gas and $\bf{5\times10^{11}}$~mol/cm$^{3}$ gas density following our chemical network. The top panel shows the temporal evolution of \CHthCN{} abundance while the bottom shows the evolution of \HCthN{} }}
\label{fig:abundance_time}
\end{figure*}

\begin{figure*}\centering
\resizebox{1.1\linewidth}{!}{\includegraphics{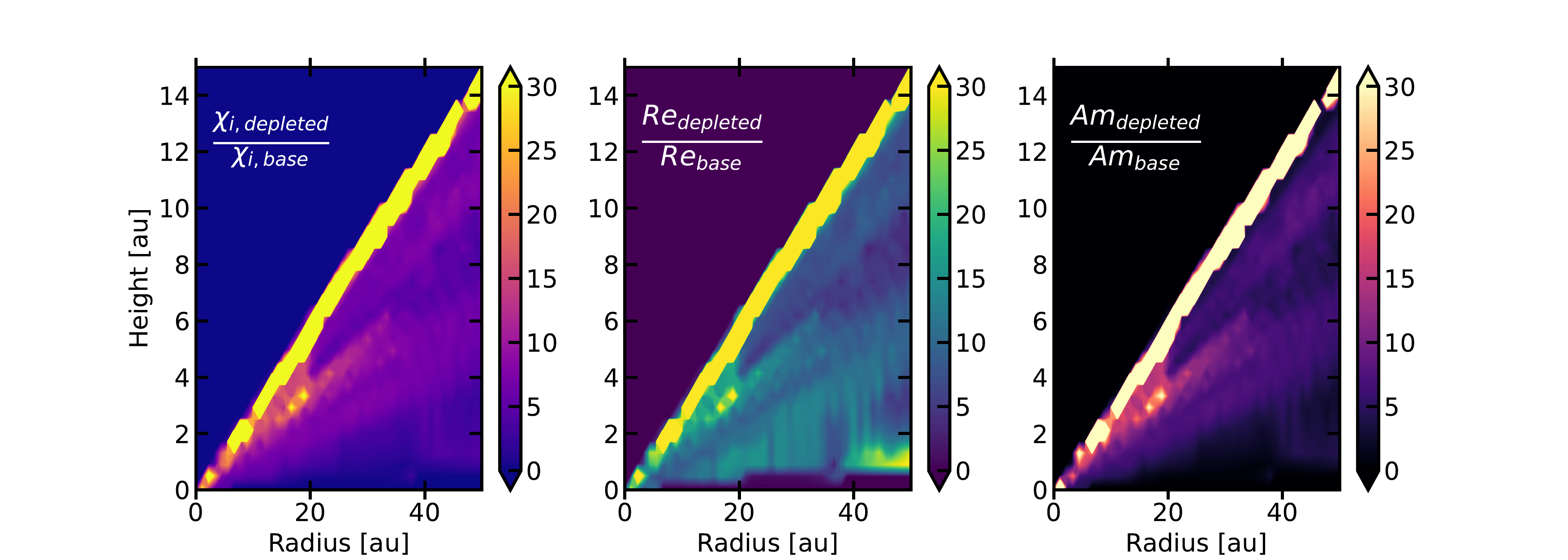}}
\caption{{ A comparison of the terms that effect the MRI strength in two distict models. The final ion abundance ($\chi_{i}$ left) calculated Reynolds number (Re, middle) and ambipolar diffusion term (Am, right) are shown in a model depleted of small dust (gas-to-dust = 5,000) versus a baseline model (gas-to-dust = 500). The depleted model corresponds to the model that reproduces observations of \CHthCN{}, HCN, and \HCthN{}, see Figure 2 of main text.}}
\label{fig:AmRe}
\end{figure*}

\begin{figure*}\centering
\resizebox{1\linewidth}{!}{\includegraphics{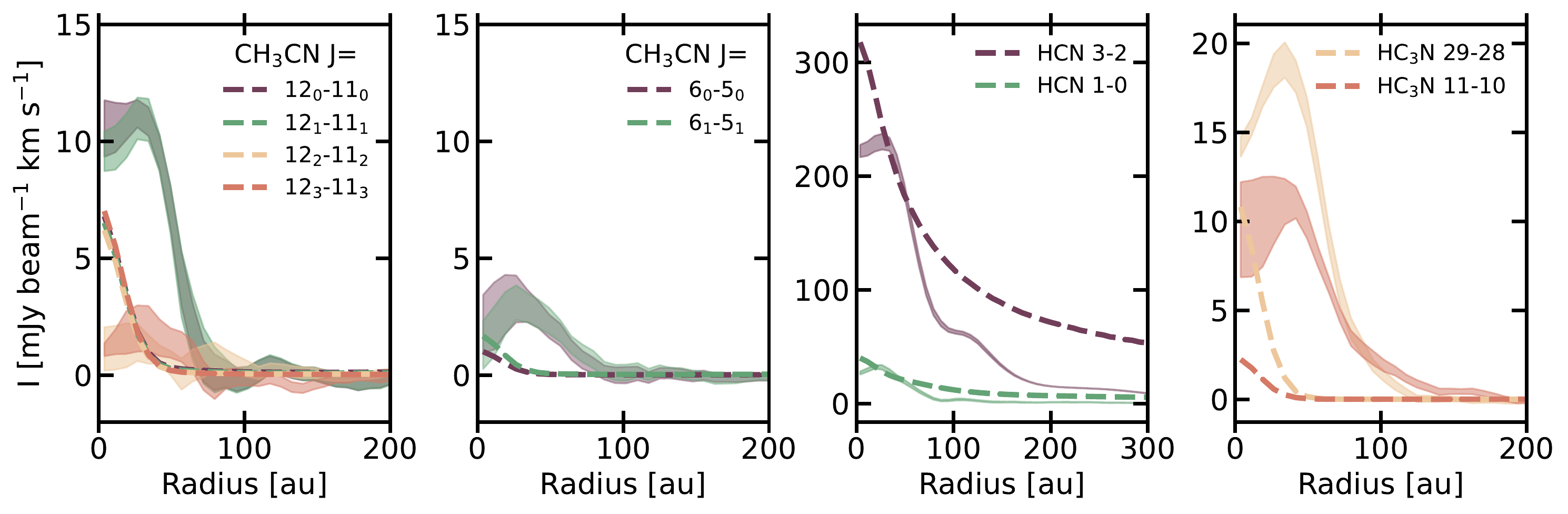}}
\caption{{ An HD 163296 model with an elevated C-to-O ratio and normal gas-to-dust ratio in the atmosphere. The model is represented by dashed lines while observations are the thick low opacity lines, where the thickness corresponds to the uncertainty in the flux. The model predicts that each K-line for \CHthCN{} J=12-11 have nearly identical morphology and intensities. \HCthN{} is predicted to have a centrally peaked radial intensity profile while observations show a plateau or central dip. By simply decreasing the total mass thus total surface density of the small dust by a factor of 10, the model is much more consistent with observations across all three molecules and their multiple transitions. }}
\label{fig:noUV_model}
\end{figure*}

\begin{figure*}\centering
\resizebox{1\linewidth}{!}{\includegraphics{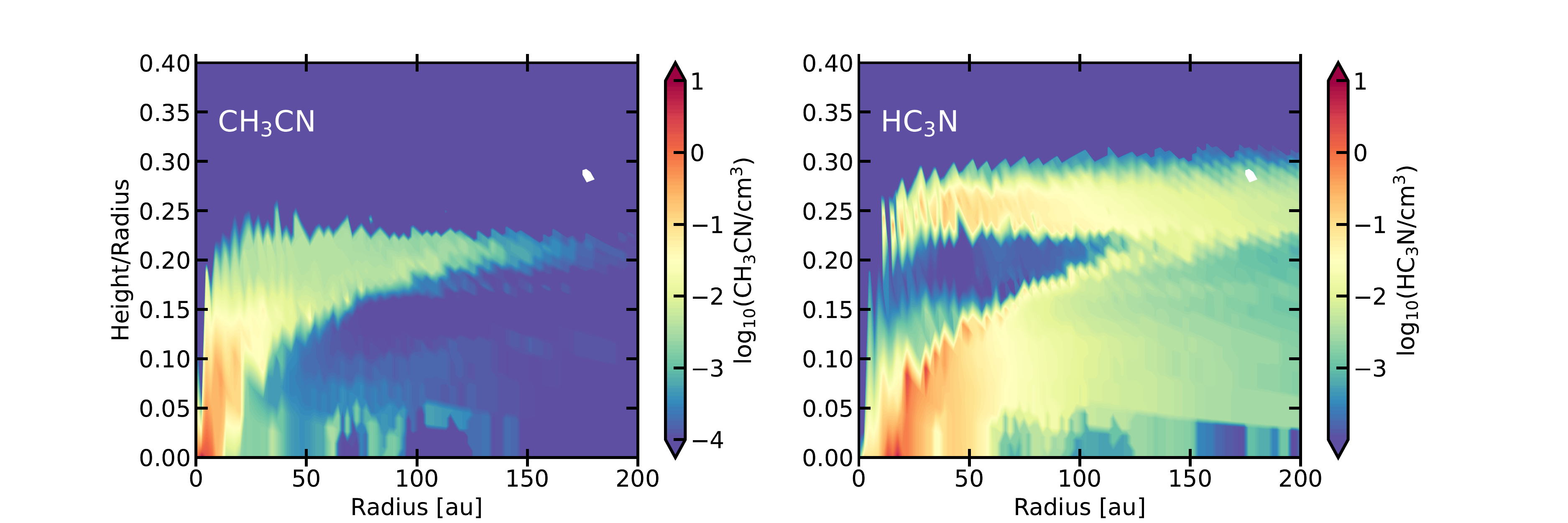}}
\caption{{ The radial and vertical number density distributions of \CHthCN{} and \HCthN{} in a model with a high C-to-O ratio  and normal gas-to-dust ratio. C-to-O is equal to 2 throughout the whole disk, and the atmospheric gas-to-dust ratio is equal to 500 (corresponding to the model results in Figure \ref{fig:noUV_model}). This is contrasted with Figure 3, where there is more organic emission deeper in the disk.   }}  
\label{fig:2d_abundances_noUV}
\end{figure*}

\begin{figure*}\centering
\resizebox{1\linewidth}{!}{\includegraphics{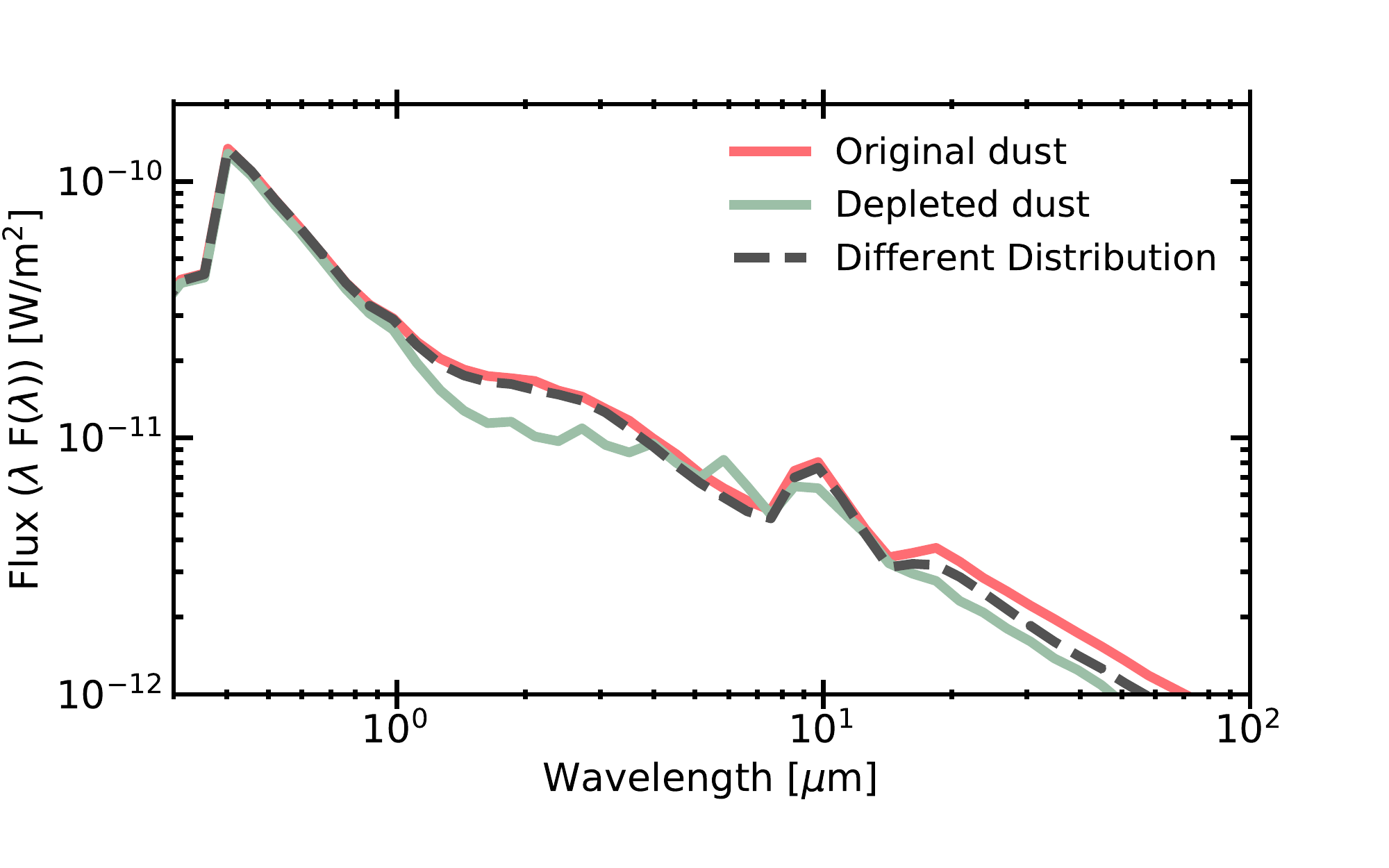}}
\caption{ Three simulated SEDs for different protoplanetary disk dust populations. The `depleted dust' model has 10 times less mass in the small dust population than the `Original dust' model. By altering the UV attenuation via increasing the minimum dust radius from 0.005~$\mu$m to 0.1~$\mu$m (dashed grey line) we reproduce the SED features and intensity as seen in the `original' model. }
\label{fig:SED_example}
\end{figure*}

\begin{acknowledgments}
\section{Acknowledgements}
Jenny Calahan is the corresponding author and can be contacted via jcalahan@umich.edu

J.K.C. acknowledges support from the National Science Foundation Graduate Research Fellowship under Grant No. DGE 1256260 and the National Aeronautics and Space Administration FINESST grant, under Grant no. 80NSSC19K1534.  E.A.B. acknowledges support from NSF AAG Grant \#1907653. A.D.B. acknowledges support from NSF AAG Grant \#1907653. E.A.R. acknowledges support from NSF AST 1830728. Support for J. H. was provided by NASA through the NASA Hubble Fellowship grant \#HST-HF2-51460.001-A awarded by the Space Telescope Science Institute, which is operated by the Association of Universities for Research in Astronomy, Inc., for NASA, under contract NAS5-26555.  This work was supported by a grant from the Simons Foundation 686302 and by an award from the Simons Foundation 321183FY19, K\"O. This material is based upon work supported by the National Science Foundation under Grant No. AST-1907832. J.D.I. acknowledges support from an STFC Ernest Rutherford Fellowship (ST/W004119/1) and a University Academic Fellowship from the University of Leeds. C.W.~acknowledges financial support from the University of Leeds, the Science and Technology Facilities Council, and UK Research and Innovation (grant numbers ST/T000287/1 and MR/T040726/1). V.V.G. gratefully acknowledges support from FONDECYT Regular 1221352, ANID BASAL projects ACE210002 and FB210003, and ANID, -- Millennium Science Initiative Program -- NCN19\_171.

We thank Tim Harries for help with and providing access to the TORUS modeling program. 

\end{acknowledgments}

\section*{Data Availability}
The data that support the findings of this study can be obtained as part of the MAPS program and are publicly available via alma-maps.info. Data regarding TW Hya can be obtained via data-rich figures in Calahan et al 2021 published on the online publication of the Astrophysical Journal article. 

This paper makes use of the following ALMA data: ADS/JAO.ALMA\#2018.1.01055.L. and 2016.1.01046.S ALMA is a partnership of ESO (representing its member states), NSF (USA) and NINS (Japan), together with NRC (Canada), MOST and ASIAA (Taiwan), and KASI (Republic of Korea), in cooperation with the Republic of Chile. The Joint ALMA Observatory is operated by ESO, AUI/NRAO and NAOJ. The National Radio Astronomy Observatory is a facility of the National Science Foundation operated under cooperative agreement by Associated Universities, Inc.

\section*{Code availability}
This study relied on the following publicly available coding packages:
rac2d: https://github.com/fjdu/rac-2d, RADMC-3D: https://www.ita.uni-heidelberg.de/~dullemond/software/radmc-3d/, and  GoFish: https://github.com/richteague/gofish. TORUS is a private code developed by Tim Harries and collaborators. 

%


\bibliography{sample}{}
\bibliographystyle{plainnat}  



\end{document}